\documentclass[sigconf]{acmart}

\AtBeginDocument{%
  \providecommand\BibTeX{{%
    \normalfont B\kern-0.5em{\scshape i\kern-0.25em b}\kern-0.8em\TeX}}}

\setcopyright{acmcopyright}
\copyrightyear{2018}
\acmYear{2018}
\acmDOI{10.1145/1122445.1122456}

\acmConference[Woodstock '18]{Woodstock '18: ACM Symposium on Neural
  Gaze Detection}{June 03--05, 2018}{Woodstock, NY}
\acmBooktitle{Woodstock '18: ACM Symposium on Neural Gaze Detection,
  June 03--05, 2018, Woodstock, NY}
\acmPrice{15.00}
\acmISBN{978-1-4503-XXXX-X/18/06}
\usepackage{multirow}

\begin{document}

\title{Represent Items by Items: An Enhanced Representation of the Target Item for Recommendation}


\author{Yinjiang Cai$^{1,2}$, Zeyu Cui$^{2,3}$, Shu Wu$^{2,3}$, Zhen Lei$^{1,2}$, Xibo Ma$^{1,2,*}$}
\affiliation{
\institution{$^1$CBSR\&NLPR, Institute of Automation, Chinese Academy of Sciences, Beijing, China}
\institution{$^2$School of Artificial Intelligence, University of Chinese Academy of Sciences, Beijing 100049, China}
\institution{$^3$Center for Research on Intelligent Perception and Computing, Institute of Automation, Chinese Academy of Sciences}
}
\email{{yinjiang.cai,zeyu.cui,shu.wu,zlei,xibo.ma}@nlpr.ia.ac.cn}
\thanks{*To whom correspondence should be addressed.}


\renewcommand{\shortauthors}{Trovato and Tobin, et al.}

\begin{abstract}
Item-based collaborative filtering (ICF) has been widely used in industrial applications such as recommender system and online advertising. It models users' preference on target items by the items they have interacted with. Recent models use methods such as attention mechanism and deep neural network to learn the user representation and scoring function more accurately. However, despite their effectiveness, such models still overlook a problem that performance of ICF methods heavily depends on the quality of item representation especially the target item representation. In fact, due to the long-tail distribution in the recommendation, most item embeddings can not represent the semantics of items accurately and thus degrade the performance of current ICF methods. In this paper, we propose an enhanced representation of the target item which distills relevant information from the co-occurrence items. We design sampling strategies to sample fix number of co-occurrence items for the sake of noise reduction and computational cost. Considering the different importance of sampled items to the target item, we apply attention mechanism to selectively adopt the semantic information of the sampled items. Our proposed Co-occurrence based Enhanced Representation model (CER) learns the scoring function by a deep neural network with the attentive user representation and fusion of raw representation and enhanced representation of target item as input. With the enhanced representation, CER has stronger representation power for the tail items compared to the state-of-the-art ICF methods. Extensive experiments on two public benchmarks demonstrate the effectiveness of CER.
\end{abstract}



\maketitle

\section{Introduction}

Personalized recommendation have been widely used in many online services such as E-commerce, advertising and social media sites. The core of a personalized recommender system is modeling the preference of users on items based on their behaviors. Various methods have been applied to boost the recommendation performance, among these methods, collaborative filtering (CF) is a milestone one.

\begin{figure}[t]
\centering
\includegraphics[width=0.46\textwidth]{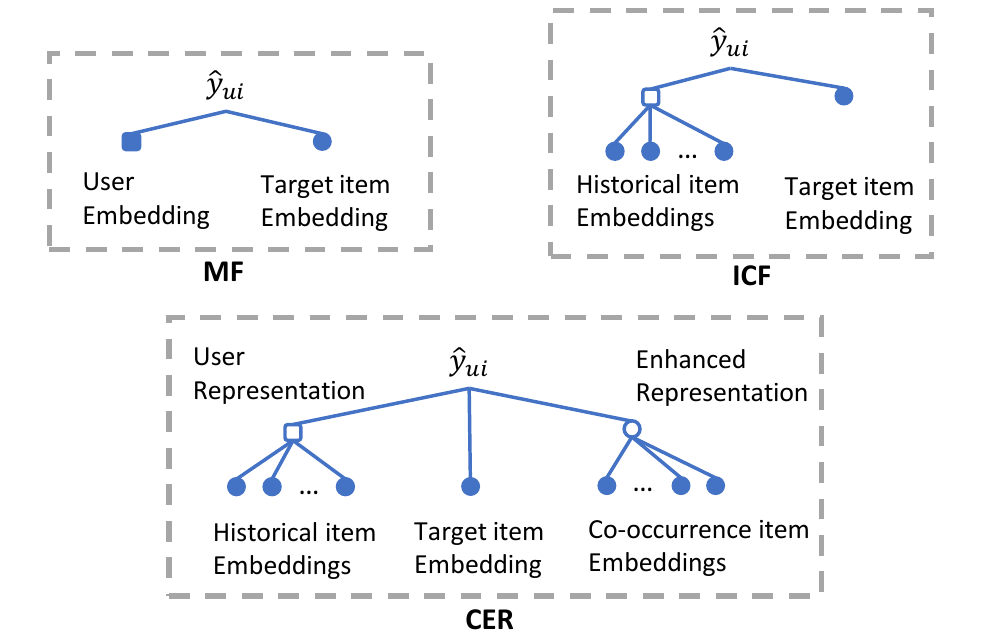} 
\caption{Illustration of different CF methods. Solid square and circle represent user and item embedding. The hollow node denotes a representation that encodes other solid nodes. From MF to our CER, ICF encodes historical items to enhance the user representation, CER encodes co-occurrence items to enhance the target item representation.}
\label{intro}
\end{figure}

A series of CF based methods focused on how to represent the user more accurately. As shown in figure \ref{intro}, Matrix Factorization (MF) \cite{mf,bpr,fastmf} represents users by the embedding of user ID, which is a user-based CF (UCF) method. In MF, interest between user and item is calculated by using the inner product of the user embedding and item embedding. Previous works \cite{fism,nais,dicf} have suggested that item-based CF (ICF) methods which directly represent the users by their historically interacted items have advantages in accuracy, real-time personalization and interpretability \cite{dicf}. Among these methods, FISM \cite{fism} predicts the interest by calculating the inner product between historical items and target item instead of between user embedding and item embedding in UCF. NAIS \cite{nais} further extends FISM by dynamically adjusting the importance of historical items according to different target items. 

\begin{figure}[t]
\centering
\includegraphics[width=0.46\textwidth]{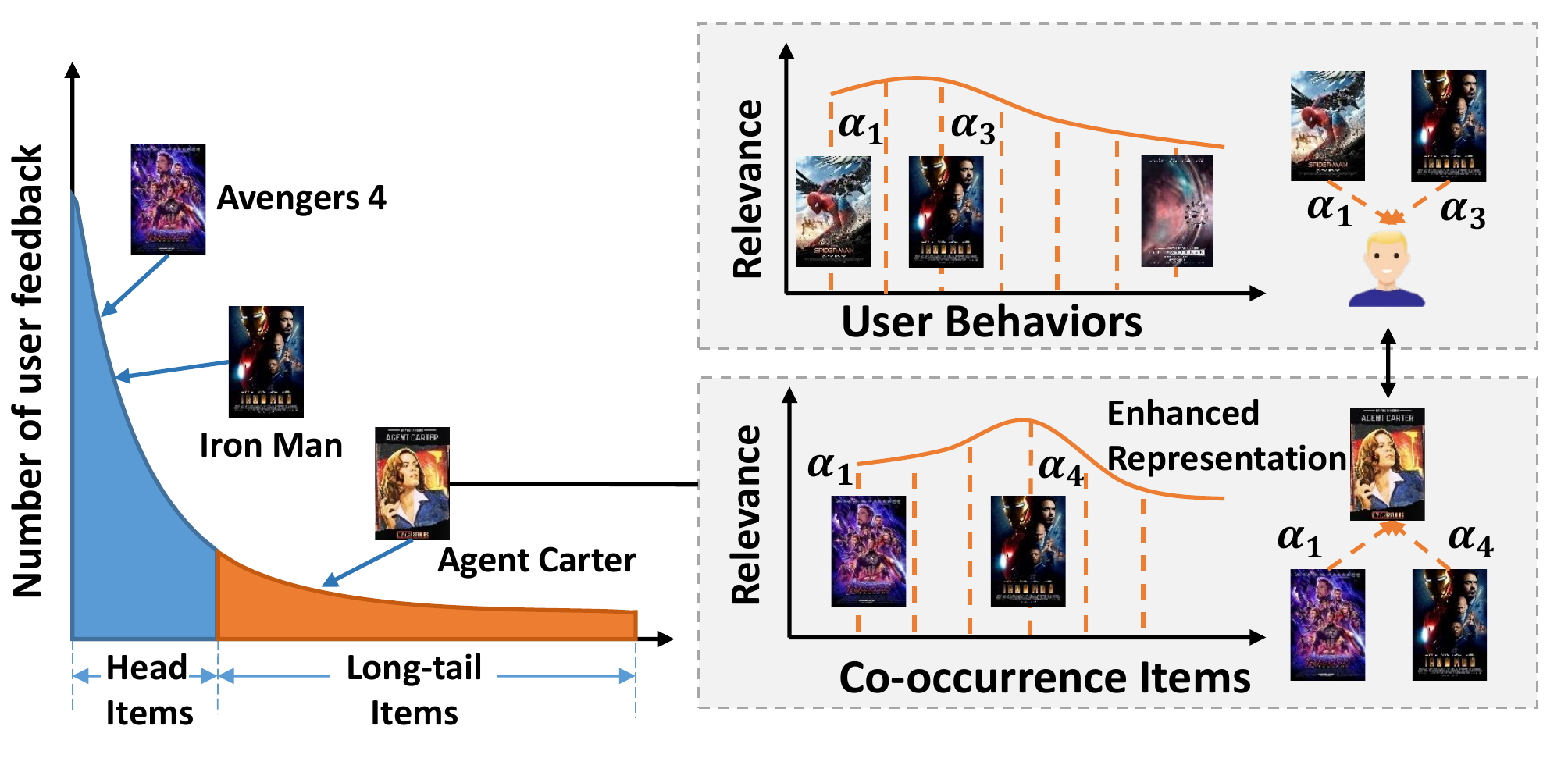} 
\caption{Illustration of long-tail distribution and enhanced representation. $\alpha$ represents the relevance between movie `Agent Carter' and other movies. For the tail item `Agent Carter', its enhanced representation are learned by the co-occurrence items such as `Avengers 4' and `Iron Man' basing on their relevance $\alpha_1$ and $\alpha_4$. Thus Marvel movie `Agent Cater' is more likely to be recommended to fans of Marvel movies.}
\label{introduction}
\end{figure}

Although these ICF methods demonstrated their performance in the recommendation, we argue that they neglected a significant issue, the quality of item embeddings. From figure \ref{intro}, for ICF methods, it is not difficult to find that if the embedding of the target item is not accurate, the final prediction will be deviated greatly. Then we will ask, is the quality of the current item embeddings good enough in fact? As we all know that the quality of one embedding depends on the training method and the frequency of the item. As the higher frequency of the item, the more fully the model learns it. Unfortunately, long-tail distributions widely exist in many real-world datasets \cite{zhang2020twotales}. As shown in the left part of figure \ref{introduction}, the long-tail distribution in the recommender system determines that a large number of items can not be trained well as they have little user feedback. Thus these tail items will finally degrade the performance of the model. To sum up, we find that the performance of the ICF methods depends heavily on the quality of the item representations, but current models can not represent most items well due to the long-tail distribution. The experiment results in figure \ref{wrong_class} show that the accuracy of the current ICF method for low-frequency items is much lower than that for high-frequency items, which verifies the problem mentioned above.


To alleviate the aforementioned issue of ICF methods, here we propose an enhanced representation to complement the semantics of the target items, which is shown in figure \ref{intro}. We all know that items which user have interacted with are out of the same user interest. Thus these items are highly complementary in semantics. Here we proposed to learn the enhanced representation of the target item based on the co-occurrence items of it. Among these co-occurrence items, head items that have high quality of embeddings can provide accurate semantics for the target item. Meanwhile, a large number of tail items among them can still provide relevant information for the target item by adopting attention mechanism. Specifically, for each target item, we search for all items which have co-appeared with the target item and construct a co-occurrence item set. For the sake of noise reduction and computational complexity, we design sampling strategies to sample fix number of items from the co-occurrence item set. In these sampling strategies, two criteria are considered. One is to sample items with high semantic similarity to the target item. Another is to sample items whose embeddings have low semantic uncertainty. Considering the different semantic relevance between co-occurrence items and target item, co-occurrence items should also contribute differently to the enhanced representation of the target item. To distinguish the different importance of co-occurrence items, here we employ attention mechanism \cite{li2017attentional,xiao2017attentional} which has been widely used in neural representation learning. Figure \ref{introduction} illustrates both the user representation and enhanced representation. With enhanced representation, our proposed model CER can also work well for large among of tail items. Extensive experiments on public datasets show that our method betters FISM for a 17 percent relative improvement in terms of NDCG@10 and 26 percent relative improvement in HR@10 on the Amazon-Books dataset.

The main contributions of this paper are summarized as follows:
\begin{itemize}

\item We highlight that the long-tail distribution limits the current expression of items and thus the poor quality of target item embedding will degrade the performance of existing ICF methods.
\item We propose an enhanced representation of the target item, a novel expression combining the idea of item-based representation with collaborative signals, which selectively learns representation from the embeddings of co-occurrence items.
\item We conduct experiments on two public datasets. Extensive results show that our model significantly outperforms the state-of-the-art methods and demonstrate its effectiveness in improving the quality of target item representation.
\end{itemize}

\section{Related Work}
Following figure \ref{intro}, we can divide each CF method into three parts, namely modeling user, modeling target item and modeling interaction. In this section, we briefly review several lines of literature closely related to these three parts.

For modeling user, early UCF methods \cite{koren2008ucf,he2017ncf} represent the user by the user ID which limits the expression of the interests of the user. Recent ICF methods aim to encode more signals in the user representation rather than only user ID in UCF. ItemKNN \cite{sarwar2001itemknn} utilizes statistical measures such as cosine similarity to estimate the similarities between items and sum the similarities between historical items and target item as the interest between user and target item. Based on the idea of ItemKNN, SLIM \cite{ning2011slim} propose to use matrix to fit the similarities between items. FISM \cite{fism} calculates the similarity between two items by the inner product of their latent vectors. NAIS \cite{nais} argues that historical items have different importance for the target item. Thus it proposed a more flexible representation of the user which uses attention mechanism to dynamically adjust the weight of each historical item. 

For modeling interaction, early methods directly use the inner product between user latent vector and item latent vector as the interaction function. Recent literature \cite{he2017deep,zhang2019deep,beutel2018deep} show that deep learning has the potential to dramatically boost the performance of recommender systems. NCF \cite{he2017ncf} proposed to learn the complex interaction function by a standard structure of MLP. DeepICF \cite{dicf} argues that the strong function learning ability of MLP is expected to capture the higher-order item relations.

For modeling target item, except for the most common use of the item ID, DMF \cite{xue2017dmf} represents an item by a multi-hot vector representing the users who have interacted with the item. DELF \cite{cheng2018delf} employs attention mechanism to discriminate the importance of the interacted users automatically. Another kind of approaches to encode items by the collaborative signal are graph-based methods \cite{he2020lightgcn,wang2019ngcf}. Graph-based methods construct an graph to represent the interactions between users and items and each vertex in this graph can be encoded by its neighbors. In nature, they still follow the idea that representing items by users, representing users by items. One of the advantages of graph-based methods is that they make it easy to capture the high-order connectivity. This is how it differs from the approaches that are neighbor-based \cite{bai2017neighbor_based,yin2012challenging} but not graph-based. In addition to the collaborative signal, some methods model items with additional side information such as visual patterns \cite{geng2015duif,chen2017acf} and knowledge \cite{zhang2016knowledge}.

\section{PRELIMINARIES}
Before we introduce our model, we firstly recapitulate the framework for standard item-based collaborative filtering. We then brief the FISM and NAIS methods as they form the basis of our method.

\subsection{Standard Item-Based CF}
The core of item-based CF is that user is no longer represented by the unique user embedding, but by the items that the user has interacted with. So the idea of item-based CF is to predict the probability that a user $u$ is interested in a target item $i$ based on the user behaviors. Formally, the prediction of ICF model can be abstracted as
\begin{equation}
    \hat{y}_{ui} = \sum_{j \in R_u}s_{ij}r_{uj},
\end{equation}
where $R_u$ denotes the set of items that the user $u$ has interacted with, $r_{uj}$ denotes the preference of user $u$ on item $j$, which can either be a binary value (implicit feedback) or a real-valued rating score (explicit feedback), $s_{ij}$ denotes the similarity between item $i$ and $j$.



\subsection{FISM and NAIS Methods}

Recent methods like FISM \cite{fism} apply inner product between latent vectors of items to express the similarity mentioned above. The prediction of FISM can be formulated as
\begin{equation}
    \hat{y}_{ui}=\frac{1}{|R_u|^\alpha}\sum_{j \in R_u}r_{uj}\mathbf{q}_i^\top \mathbf{q}_j
\end{equation}
where $\mathbf{q_i} \in \mathbb{R}^k$ and $\mathbf{q}_j \in \mathbb{R}^k$ denote the latent vectors for the target item $i$ and historical item $j$, all latent vectors are trainable parameters with embedding size $k$. Hyper-parameter $\alpha$ controls the normalization on users of different behavior length, $\alpha=0$ means the prediction is the sum over item set, $\alpha=1$ means the prediction is an average value. 

FISM assumes that each historical item contributes equally to the final prediction score, thus fails to accurately capture user interest with respect to the target item. NAIS \cite{nais} proposed a more reasonable strategy that applying dynamic weights to historical items which vary over different target items. The prediction of NAIS is formulated as
\begin{equation}
    \hat{y}_{ui}=\sum_{j \in R_u}a_{ij}\mathbf{q}_i^\top \mathbf{q}_j
\end{equation}
where $a_{ij}$ denotes the attentive weight of similarity between item $i$ and $j$. In NAIS, the attentive weight is calculated by a neural network whose inputs are the embeddings $\mathbf{q_i}$ and $\mathbf{q_j}$. 

\begin{figure*}[t]
\centering
\includegraphics[width=0.9\textwidth]{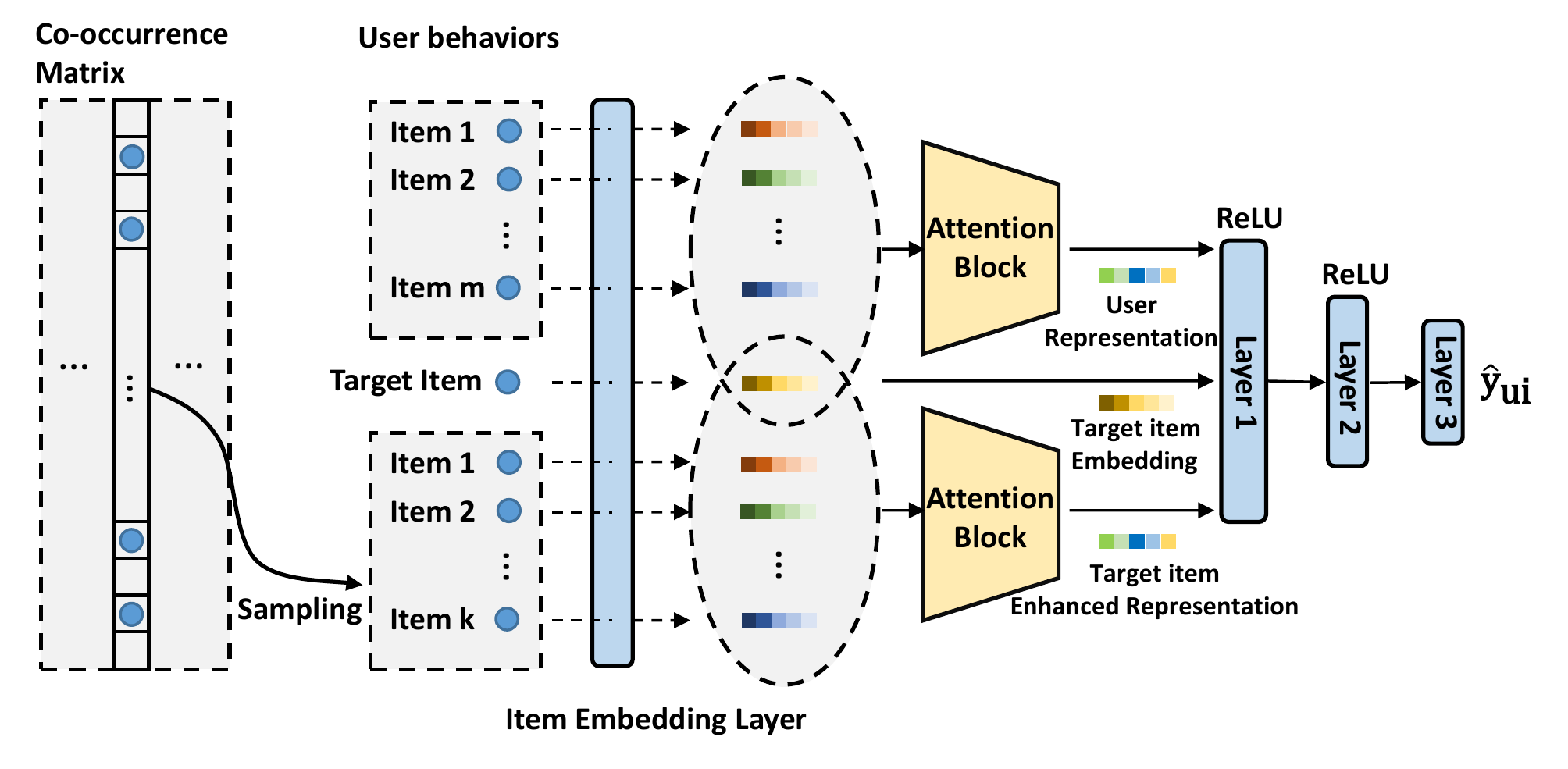} 
\caption{Overview of CER. The user representation encodes the user behaviors. The enhanced representation is learned based on the sampled co-occurrence items and the target item. Attention block uses the target item to selectively adopt information from user behaviors and co-occurrence items respectively. The neural network outputs $\hat{y}_{ui}$ as the preference of user $u$ on target item $i$.}
\label{fig1}
\end{figure*}

\section{Method}

In this section, we introduce CER in detail. First, we elaborate the design of enhanced representation of the target item. Then we introduce the sampling methods we designed. After that, we describe how to merge the raw and enhanced representations and brief the inference process of our model as shown in figure \ref{fig1}. Last, we discuss the time complexity of our model.

\subsection{Enhanced Representation}
In the previous section, we demonstrate several modeling approaches for target item using the collaborative signal. Here we follow the idea of representing items by users to introduce our methodology. 

In recommendation system, like the user can be represented by the interacted items, item can also be represented by the interacted users. For the item $i$ and the corresponding interacted user set $U_i$, the collaborative signal means that the users in this set must share part of interests with respect to the item $i$. Thus one feasible way to represent the item based on the user is to sum over the embeddings of these users. This method can be formulated as
\begin{equation}
    \mathbf{r}_i = \frac{1}{|U_i|^\alpha}\sum_{j \in U_i}\mathbf{u}_j
\end{equation}
where $\mathbf{u}_j$ denotes the embedding of user $j$, $\mathbf{r}_i$ denotes the enhanced representation of item $i$. However, each user may only have part of interests that are aligned with the target item. For instance, one user may be enthusiastic about the target item, thus his representation is highly related to the semantic information of the target item. On the contrary, one user may only have interacted with the item once for some reasons and his interest has nothing to do with the item. From this point, it is not appropriate that each user contributes equally to the $r_i$.

To address the above problem, we adopt the attention mechanism to dynamically adjust the weight of each user. The new representation can be formulated as
\begin{equation}\label{att_user}
    \mathbf{r}_i = \sum_{j \in U_i}f(\mathbf{q}_i,\mathbf{u}_j)\mathbf{u}_j
\end{equation}
where $f(\mathbf{q}_i,\mathbf{u}_j)$ denotes the function calculating the attentive weight of user embedding $\mathbf{u}_j$. The attention function can be MLP or inner product with item embedding $\mathbf{q}_i$ and user embedding $\mathbf{u}_j$ as input. In this way, user embeddings associated with the target item $i$ will dominate in the representation $\mathbf{r}_i$ and therefore $\mathbf{r}_i$ will be more accurate than the former approach.

In the above section, we introduced the disadvantages of UCF. We find that equation \ref{att_user} also has these limitations as it uses the user embedding $\mathbf{u}_j$. So we introduce an item-based representation of the user to tackle these problems. We firstly formulate the user representation $\mathbf{u}_j$ as
\begin{equation}\label{user}
    \mathbf{u}_j = \sum_{k \in R_j}g(\mathbf{q}_i, \mathbf{q}_k)\mathbf{q}_k 
\end{equation}
where the function $g(\mathbf{q}_i, \mathbf{q}_k)$ outputs the similarity between target item $i$ and historical item $k$. Some readers may find that this formulation is exactly the user representation used in both NAIS and DIN \cite{zhou2018din} (a widely used CTR model). This formulation focuses on the interest of the user with respect to the target item $i$, thus filters out some of the noise at the user level. On this basis, we further modify the formulation of enhanced representation as
\begin{equation}
    \mathbf{r}_i = \sum_{j \in U_i}f(\mathbf{q_i}, \sum_{k \in R_j}g(\mathbf{q}_i, \mathbf{q}_k)\mathbf{q}_k) \sum_{k \in R_j}g(\mathbf{q}_i, \mathbf{q}_k)\mathbf{q}_k 
\end{equation}
Actually, since the function $f$ and $g$ output real numbers, we can simplify the formulation as:
\begin{equation}
\mathbf{r}_i = \sum_{k \in O_i} ( \sum_{n \in U_k} \sum_{m \in R_n} \alpha_n \beta_m ) \mathbf{q}_k
\end{equation}
where $O_i$ denotes the co-occurrence items set of target item $i$, $\alpha$ denotes the attentive weight of the user, $\beta$ denotes the attentive weight of the historical item. In nature, this formulation aims to allocate an appropriate weight of each co-occurrence item based on two attentive weights. However, we argue that these attention mechanisms add additional constraints on learning the relevance between items. Since the part in parentheses essentially tries to calculate a correlation, we can modify it into a more flexible form:
\begin{equation}\label{enhanced}
    \mathbf{r}_i = \sum_{k \in O_i} f(\mathbf{q}_i,\mathbf{q}_k)\mathbf{q}_k
\end{equation}
where $f$ denotes the attentive function. We choose the MLP as the attentive function in this paper and introduce the details of it in the latter subsection. This formulation also meets the idea of item2vec \cite{barkan2016item2vec} that treating the co-occurrence items as items having similar semantics.

The current enhanced representation seems good as it has sufficient motivation and a concise form that demonstrates its effectiveness. However, there is still a problem hidden in the set $O_i$.

\subsection{Sampling Strategy}
Generally, the average length of the co-occurrence item set tends to be much larger than the average length of user behaviors. We analyze the datasets used later and find that these two lengths can sometimes differ by two orders of magnitude. For this kind of item set, even if adopting an efficient method, the calculation cost is still unacceptable. In addition, too large a co-occurrence item set will exacerbate the noise problem in enhanced representation. 

To alleviate the above limitations, here we design sampling strategies to sample fix number of items from the corresponding co-occurrence item set. These sampling strategies follow two criteria: 1) the semantics of the picked item is as relevant as possible to the ground truth meaning of the target item. 2) the semantic inaccuracy of picked item is as small as possible. As the enhanced representation $\mathbf{r}_i$ is based on these picked items, if the picked items have these two advantages, $\mathbf{r}_i$ will also share the same advantages.

\textbf{Global Sampling} is a sampling method that aims to decrease semantic inaccuracy of representation $\mathbf{r}_i$. We already know that the semantic accuracy is highly related to the frequency of item. Thus we define the sampling probabilities of items as
\begin{equation}
    s_k = \frac{f^g_k}{\sum_{j \in O_i}f^g_j}
\end{equation}
where $f^g_k$ denotes the times item $k$ appears in the whole dataset. In simple terms, this method tends to pick out the items whose embeddings have high accuracy.

\textbf{Local Sampling} is a sampling method that aims to increase the semantic correlation between the picked items and the target item. Obviously, we need to pick out the items that may be more relevant to the target item $i$. However, it is hard to measure the semantic correlation accurately without any side information. Here we follow the basic idea that correlation between items is proportional to their co-occurrence frequency. Thus the sampling probability can be formulated as
\begin{equation}
    s_k = \frac{f^l_k}{\sum_{j \in O_i}f^l_j}
\end{equation}
where $f^l_k$ denotes the times item $k$ appears in the co-occurrence set. Different from the global sampling, this method focuses on relevance between item $j$ and target item $i$, thus the meaning of enhanced representation is more close to the ground truth.

\textbf{Weighted Sampling} combines the global and local sampling by a weighted sum to strike a balance between correlation and inaccuracy. The sampling probability can be formulated as
\begin{equation}
    s_k = \alpha * s_k^g+ (1-\alpha)s_k^l
\end{equation}
where $s_k^g$ denotes the global sampling probability of item $k$ and $s_k^l$ denotes the local sampling probability, $\alpha$ is a hyper-parameter to control the proportion of these two methods in the final probability.

\textbf{Uniform Sampling} is an unbiased sampling strategy in which each item has the same sampling probability
\begin{equation}
    s_k = \frac{1}{|O_i|}
\end{equation}
This method can be treated as a baseline sampling strategy compared with others.

\subsection{Inference}
In this subsection, we first describe how to merge the raw representation and enhanced representation of target item. At the same time, we introduce the process of inference based on user representation and merged representation. Then we brief the design of user representation and attention mechanism used in this paper.

In the inference stage, previous works adopted inner product or MLP to calculate the interest $\hat{y}_{ui}$ between user $u$ and target item $i$. Now we get two representations of the target item, raw embedding $\mathbf{q}_i$ and enhanced representation $\mathbf{r}_i$. Intuitively, the final prediction can be formulated as
\begin{equation}
    \hat{y}_{ui} = f(\mathbf{q}_i,\mathbf{e}_u) + f(\mathbf{r}_i,\mathbf{e}_u)
\end{equation}
where $\mathbf{e}_u$ is the representation of user $u$ calculated by equation \ref{user}, $f$ is a function which can be either inner product or MLP. However, this formulation sets the weights of raw and enhanced representations as the same by default, which has a limitation in dynamically adjusting the weights of different target items. For instance, the raw representation of a head item is good enough to be used in prediction, but its co-occurrence items may not reflect its semantic information as it appears too many times thus any item (especially the popular ones) can be added into the co-occurrence item set. Therefore, the final enhanced representation will be deviated from the raw meaning of the target item and exacerbates the prediction. On the contrary, for tail items, enhanced representations can alleviate the deficiency of raw embedding in expressing semantic information, thus deserve larger weight compared with the head items.

To tackle the above problem, we apply MLP to solve the fusion of two representations and the selection of prediction function. Specifically, we directly concatenate these three vectors and use the concatenated vector as the input of MLP. The formulation can be formulated as
\begin{equation}
\begin{split}
\mathbf{c}_0 &= concat(\mathbf{q}_i,\mathbf{r}_i,\mathbf{e}_u)\\
\mathbf{c}_k &= ReLU(\mathbf{W}_k\mathbf{c}_{k-1}+\mathbf{b}_k)\\
\hat{y}_{ui} &= \mathbf{W}_K\mathbf{c}_{K-1}+\mathbf{b}_K
\end{split}
\end{equation}
where $k$ denotes the number of the layers in MLP, $K$ denotes the number of final layer in MLP, $W_k$ and $b_k$ are respectively the weight matrix and bias vector which are trainable parameters. In this design, the weights between $\mathbf{q}_i$ and $\mathbf{r_i}$ are learnable and vary over different target item $i$, which ensures that the contributions of two representations are optimal. 

Finally, we brief the design of user representation $\mathbf{e}_u$ and attention mechanism used in this paper. For user representation $\mathbf{e}_u$, we follow the previous works \cite{nais,zhou2018din} and formulate the $\mathbf{e_u}$ as
\begin{equation}
    \mathbf{e}_u = \sum_{j \in R_u}a_{ij}\mathbf{q}_j
\end{equation}
where $\mathbf{q}_j$ denotes the embedding of interacted item $j$, $a_{ij}$ denotes the weight calculated by attentive function with embedding of target item $i$ and interacted item $j$ as the input. In this work, all attentive weights (between interacted items or co-occurrence items) adopt the MLP as the function like
\begin{equation}
    \begin{split}
        a_0 &= concat(\mathbf{q}_i,\mathbf{q}_j)\\
        a_k &= ReLU(\mathbf{W_k}a_{k-1}+\mathbf{b}_{k-1})\\
        a_{ij} &= \mathbf{W}_Ka_{K-1}+\mathbf{b}_K
    \end{split}
\end{equation}
where $\mathbf{W}_k$, $\mathbf{b}_k$ and $a_k$ denote the weight matrix, bias vector and output vector of the $k$th hidden layer, respectively. We will report the detailed setting of both prediction MLP and attentive MLP in the experiment section.

\subsection{Time complexity}
In this subsection, we discuss the time complexity of our model. In the training stage, we first show that the time complexity of FISM in evaluating a prediction is $O(d|R_u|)$, where $d$ denotes the size of embedding. For the NAIS, as reported in reference \cite{nais}, the time complexity is $O(d'd|R_u|)$, where $d'$ denotes the attention factor. For our model, the time complexity in calculating the attentive function is $O(\sum_{k=1}^Kd_{k-1}d_k)$, thus the final time complexity of our model is $O(|R_u|\sum_{k=1}^Kd_{k-1}d_k + |O_i|\sum_{k=1}^Kd_{k-1}d_k + \sum_{k=1}^Kd_{k-1}'d_k')$, where three parts denote the time complexity of calculating user representation, enhanced representation and prediction respectively. $d_k'$ denotes the size of $k$th hidden layer in prediction MLP.

The time complexity of our model in the training stage seems pretty large compared with other methods. Actually, if we look back at the equation \ref{enhanced}, we will find that the enhanced representation has nothing to do with the users and only depends on the sampled co-occurrence items. Therefore, once the training of the model is completed, the enhanced representation of items can be calculated offline. We can first calculate the enhanced embedding matrix and then store it as part of the model. In the stage of online inference, the model can directly look up the enhanced embedding matrix and skip the calculation of it. So the time complexity in testing stage actually is $O(|R_u|\sum_{k=1}^Kd_{k-1}d_k + \sum_{k=1}^Kd_{k-1}'d_k')$.

\section{EXPERIMENTS}
In this section, we conduct experiments on two public datasets with the aim of answering the following research questions:

\begin{itemize}
\item \textbf{RQ1:} How does our proposed model perform compared to other state-of-the-art recommender models?
\item \textbf{RQ2:} How does our proposed model perform on target items with low frequency?
\item \textbf{RQ3:} How does the choice of sampling strategy impose an influence on the performance of our model?
\item \textbf{RQ4:} Besides the performance on tail items, does the enhanced representations actually better than the embeddings for tail items? Could we see the differences visually? 
\end{itemize}


\begin{table}[t]
\caption{The statistics of datasets.}
\label{dataset}
\centering
\begin{tabular}{l r r r r}
\toprule
Dataset & \#Users & \#Items & \#Interactions & Density \\
\midrule
Books       & 603,668   & 367,982   & 8,898,041  & 0.004\% \\





Beauty      & 22,363    & 12,101    & 198,502   & 0.073\%\\


\bottomrule
\end{tabular}
\end{table}

\begin{table*}[t]
\caption{Model comparison on public datasets. Relative Improvement (RI) is based on FISM.}
\centering
\label{results}

\begin{tabular}{ l  crcr crcr}
\toprule
\multirow{2}{*}{Methods} & \multicolumn{4}{c}{Books} & \multicolumn{4}{c}{Beauty}\\
\cmidrule(lr){2-5}
\cmidrule(lr){6-9}
&HR& \multicolumn{1}{c}{RI} &NDCG& \multicolumn{1}{c}{RI} &HR& \multicolumn{1}{c}{RI} &NDCG& \multicolumn{1}{c}{RI}\\

\midrule
MF          &0.4910 &-27.54\%    &0.3169 &-33.74\%  
            &0.3628 &-19.36\%   &0.2074 &-25.13\%   \\
MLP         &0.4917 &-27.44\%    &0.3171 &-33.70\%  
            &0.3777 &-16.05\%   &0.2101 &-24.15\%     \\
Youtube Rec &0.7162 &5.70\%    &0.4991 &4.35\%  
            &0.4399 &-2.22\%    &0.2588 &-6.57\%    \\
FISM        &0.6776 &0.00\%     &0.4783 &0.00\%  
            &0.4499 &0.00\%     &0.2770 &0.00\%     \\
NAIS        &0.6774 &-0.03\%    &0.4824 &0.86\%  
            &\underline{0.4617} &\underline{2.62\%}    &\underline{0.2846} &\underline{2.74\%}   \\
DIN         &\underline{0.7193} &\underline{6.15\%}    &\underline{0.5069} &\underline{5.98\%}  
            &0.4482 &-0.38\%    &0.2763 &-0.25\%    \\
CER        &\textbf{0.7987} &\textbf{17.87\%}     &\textbf{0.6062} &\textbf{26.74\%}  
            &\textbf{0.4968} &\textbf{10.42\%}     &\textbf{0.3061} &\textbf{10.51\%}     \\
\bottomrule
\end{tabular}
\end{table*}

\subsection{Experimental Settings}
\subsubsection{Dataset Description}
To evaluate the effectiveness of our model, we conduct experiments on two real-world datasets: Amazon-Beauty and Amazon-Books, which vary in terms of size, domain and sparsity. Table \ref{dataset} summarizes the statistics of the two datasets.

Amazon dataset \cite{amazon} is widely used as a benchmark dataset in the recommendation. Here we choose two subsets of Amazon product dataset to conduct our experiments, containing Beauty and Books. These datasets are all reduced to satisfy the 5-core property.

All datasets are truncated at length 200 (keep the last 200 behaviors of users). Given all behaviors of a user $[x_1,x_2,\dots,x_n]$, we construct training set by taking the sequence $[x_1,\dots,x_{n-2}]$ as user behaviors and the task is to predict the (n-1)-th item. In the testing stage, the task is to predict the n-th item based on the behaviors of user $[x_1,\dots,x_{n-1}]$. To tune hyper-parameters, we select half of the users as the validation set and treat the remaining users as the testing set. For each interaction, we pair it with one negative item that the user did not interact with before.

\subsubsection{Evaluation Metrics.}
For easy and fair evaluation, following references \cite{bayer2017evaluation,hornik1989evaluation}, we randomly sample 100 negative items for each positive item and rank the test item among the 100 items for each user. We employ evaluation metrics including Hit Ratio (HR) \cite{deshpande2004hr} and Normalized Discounted Cumulative Gain (NDCG) \cite{he2015ndcg} at position 10. Considering we have only one positive item for each user, HR@K is equivalent to Recall@K and proportional to Precision@K. All these metrics are that the higher the values are, the better the performance is. All experiments are repeated 5 times on the testing set and averaged values are reported.

\subsubsection{Compared Methods}
We choose several baselines to evaluate the effectiveness of our model.
\begin{itemize}
    \item {\bf MF.} This is a standard matrix factorization that exploits the user-item direct interactions. 
    \item {\bf MLP.} This method replaces the inner product in MF with a multi-layer perceptron to learn the scoring function about user and item embeddings. Here we adopt the same structure of MLP used in our prediction stage, which has 3 layers and optimizes the model with the pointwise log loss.
    \item {\bf Youtube Rec \cite{covington2016youtube}.} YouTube employed this method in their video recommendation. This method averages the embeddings of videos the user has interacted with and feeds it into a feed-forward neural network. 
    \item {\bf FISM.} FISM is one of the state-of-the-art item-based CF methods. It represents users by their interacted items and adopts the inner product as the scoring function.
    \item {\bf NAIS.} NAIS additionally introduces attention to adjust the representation of users dynamically for different target items. Here we choose the version which uses the element-wise product based attention function.
    \item {\bf DIN.} This method is a widely used CTR prediction model. It shares a similar structure to NAIS, but it adopts the MLP as the scoring function.
\end{itemize}

\begin{figure*}[t]
\centering
\includegraphics[width=1.0\textwidth]{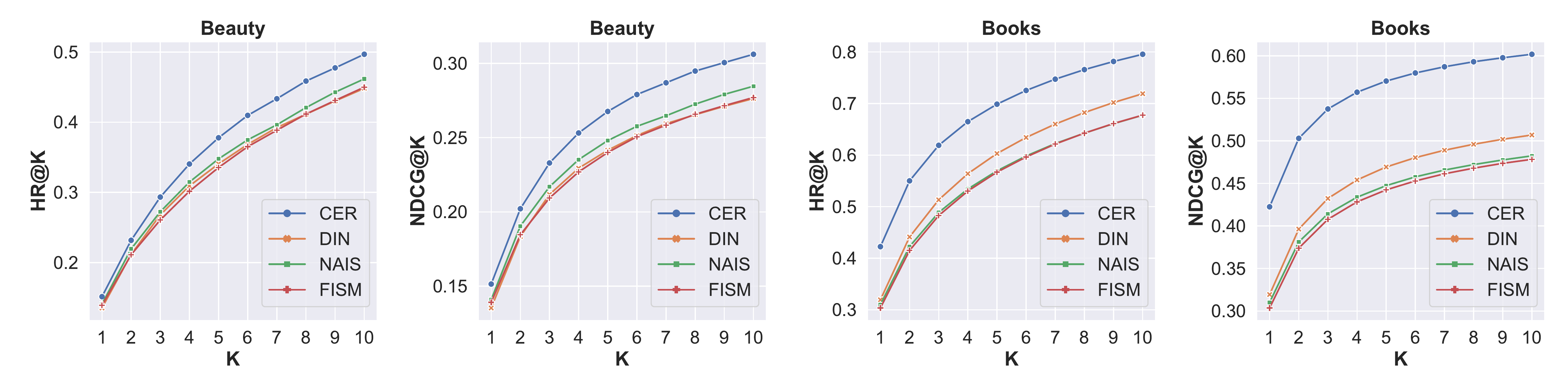} 
\caption{Performance of Top-K recommendation with ranking positions K ranging from 1 to 10.}
\label{topk}
\end{figure*}

\subsubsection{Parameter Settings}
We implement our model and all baselines in Tensorflow. For the hyper-parameters, we apply a grid search on the validation set: the embedding size is tuned among \{8, 16, 32, 64\}, the learning rate in \{0.03,0.1,0.3,1.0\}. For the hyper-parameter $\alpha$ in the model FISM and NAIS, the range is \{0, 0.1, ..., 0.9,1\}. The scaled factor $\beta$ of attention in NAIS is set as 0.5 which is the default setting in their paper. For all models, we set the batch size as 32 and choose the SGD as the optimizer. Both the prediction MLP and attentive MLP used in our model adopt the same structure that the sizes of hidden layers are [80, 40, 1] with ReLU as the active function. For the hyper-parameter $\alpha$ we used in weighted sampling, it was set to 0.6 for the Beauty dataset and 0.2 for the Books dataset.

\subsection{Performance Comparison (RQ1)}
In this subsection, we compare our proposed model with other baselines. Table \ref{results} reports the performance of HR@10 and NDCG@10 of all compared methods. 

First, we can see that CER achieves the best performance on both datasets and surpasses the state-of-the-art methods NAIS and DIN by a large margin. On the Beauty dataset, our proposed method outperforms FISM with about 10.42\% relative improvement in terms of HR and 10.51\% relative improvement in terms of NDCG. On the Books dataset, the relative improvement is 17.87\% HR and 26.74\% NDCG respectively. Basically, we can divide the baselines into three categories. First, UCF methods including MF and MLP. Second, representing User By Items (UBI) methods including Youtube, FISM, NAIS and DIN. Last, we representing both user and item by items method. As shown in table \ref{results}, among each type of methods, different tricks like applying MLP or attention lead to a certain amount of difference in performance. However, the performance of methods from different types tends to show an obvious gap. 

For UCF methods, whether the inner product or MLP is adopted as the interaction function, it always achieves the worst performance among all methods. This is because representing the interests of user by a single embedding limits the ability of the model to fit and generalize. 

Compared with UCF methods, UBI methods achieve great improvements on both datasets. These improvements are attributed to that representing users with their consumed items encodes more information than UCF methods, which improves the quality of user expression. Among these methods, the performance of methods that apply no attention like Youtube Rec and FISM is slightly lower than the performance of NAIS and DIN which are attention-based methods. It demonstrates that dynamical adjusting the expression of user interest with respect to target items helps improve the accuracy of user modeling. 

Based on the UBI methods, our proposed CER provides an item-based representation of the target item based on its co-occurrence items. This enhanced representation contributes to alleviating the semantic loss of tail items and thus improve the performance of ICF methods, especially for low-frequency target items. As we can see in table \ref{results}, our method significantly outperforms all UBI methods on both datasets, which demonstrates the effectiveness of our method.

Figure \ref{topk} shows the performance of Top-K recommendations with ranking position K ranging from 1 to 10. As can be seen, CER outperforms UBI methods consistently across positions. Among all baselines, attention-based methods NAIS and DIN perform well on Beauty and Books dataset respectively.

\begin{figure*}[t]
\centering
\includegraphics[width=1.0\textwidth]{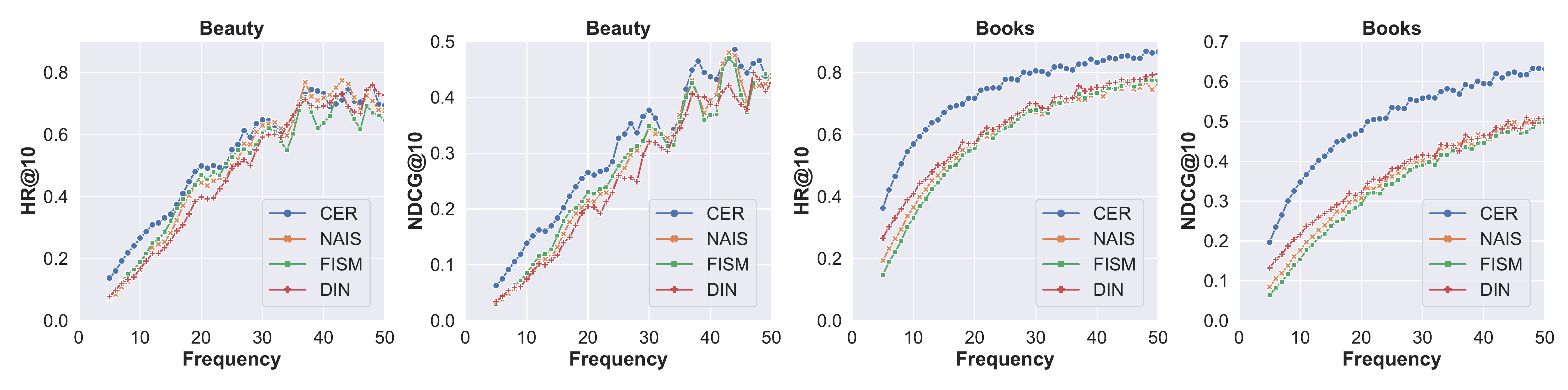} 
\caption{Performance comparison on low frequency target items}
\label{wrong_class}
\end{figure*}

\begin{figure*}[t]
\centering
\includegraphics[width=1.0\textwidth]{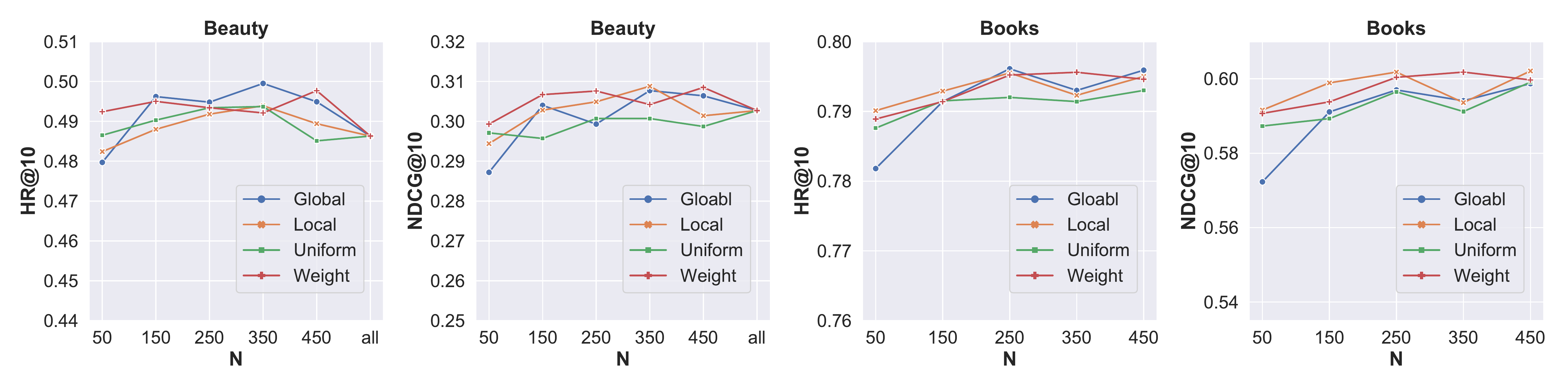} 
\caption{Performance of different sampling strategies. 'all' means directly using all co-occurrence items without sampling.}
\label{ablation}
\end{figure*}

\subsection{Performance on Low Frequency Target Items (RQ2)}
To figure out how does our proposed model performs on low frequency target items, we conduct experiments to demonstrate the effectiveness of our model, especially for the tail target items. Here we split the target items in the testing set by their global frequencies. For each frequency, we calculate both evaluation metrics based on the items with that frequency. Here we select the frequencies to range from 0 to 50 for purpose of focusing on low frequency target items.

Figure \ref{wrong_class} shows the performance on different frequencies of items. Firstly, as we can see, in all UBI methods, the values of both HR and NDCG increase with the increase of target item frequency. It proves that the frequency of the target item does have a powerful influence on the performance of UBI methods. In addition, we can see that the low frequency target items (frequencies ranging from 0 to 30) perform far below the overall level of the model. Since the long-tail distribution, substantial such items can dramatically drag down the overall performance of the model.

Secondly, on the Beauty dataset, our proposed model outperforms other UBI methods, especially for frequencies ranging from 0 to 30. On the Books dataset, our method is far superior to other UBI approaches on both evaluation metrics. This demonstrates that enhanced representation is effective in boosting the performance of UBI model on the low frequency target item. To illustrate figure \ref{wrong_class} more clearly, we take the Books dataset as an example. Considering the HR@10 value when the frequency equals 10, the best performance of UBI methods is DIN which is almost 0.41. However, our proposed model achieves almost 0.57 on the same frequency which has a nearly 39 percent relative improvement on the basis of DIN. Considering the corresponding frequency when NDCG@10 equals 0.5, the best performing UBI method DIN has a frequency of almost 44. For our proposed model, the corresponding frequency is almost 21. That is to say, to achieve the same performance on NDCG@10, our method only need the target item to appear 21 times in the dataset and the UBI method need at least 44 times. In general, the results show that our proposed model has incomparable advantages over other ICF models in low frequency target items.

\begin{figure*}[t]
\centering
\includegraphics[width=1.0\textwidth]{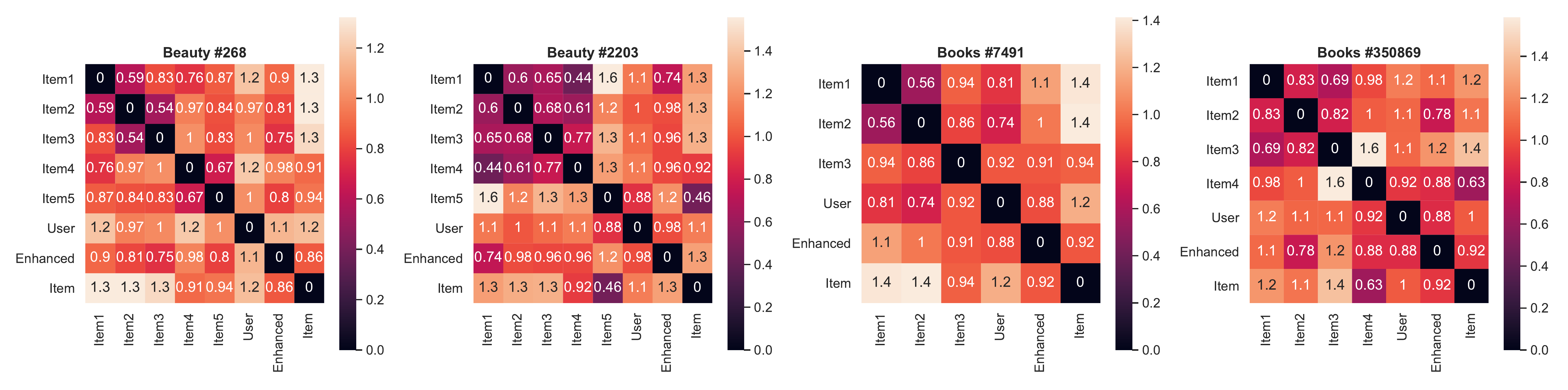} 
\caption{Visualization of cosine distances between representations including user representation, enhanced representation, target item embedding and embeddings of historical items with the same category of target item (Item1-Item5). The left two users \#268 and \#2203 are sampled from the Beauty dataset. Another two users \#7491 and \#350869 are sampled from Books dataset.}
\label{visual}
\end{figure*}

\subsection{Performance of Different Sampling Strategies (RQ3)}
Considering noise reduction and computational efficiency, we design several sampling strategies to sample fix number of items from the co-occurrence item set, following different criteria. The question is does the choice of sampling strategy play an important role in our proposed model? To figure out this question, we conduct experiments to show the influence of different sampling strategies and the influence of the number of the sampled items. We choose the four aforementioned sampling methods and the numbers of sampling include [50, 150, 250, 350, 450]. For Beauty dataset, we additionally add an experiment that directly using all co-occurrence items without sampling. Considering the computational cost, this experiment was not performed on the Books dataset.

Figure \ref{ablation} shows the experiment results of different sampling strategies. As we can see, the uniform sampling method is basically inferior to other sampling methods in most datasets and evaluation metrics. It is because that uniform sampling is an unbiased sampling method that ignores some important information such as global and local frequency. For other sampling methods, the global sampling method works well on HR but has relatively bad performance on NDCG. Local sampling, in contrast, performs well on NDCG for the most parts, but not so well on HR. This may be because that different concerns of the sampling method have different effects on the evaluation metrics. For instance, local sampling tends to pick out items that have high semantic similarity, thus it performs better on NDCG which emphasizes more on accuracy. As the weighted sampling balances the global and local sampling and it works well in most cases, here we recommend weighted sampling as a default choice.

Actually, for the numbers of sampled items, as long as $N$ is not too small say 50 or less, it does not make much influence on the model with the weighted sampling method. As shown in figure \ref{ablation}, the important thing is that taking all co-occurrence items into account may not contribute to the performance of the model. In fact, we can directly use the length of user truncation $M$ (in this paper $M=200$) as a reference, say $N= kM$ where $k$ is a real number. In this setting, a basic guarantee can be obtained is that the computational cost is acceptable and the performance is good enough.

\subsection{Visualization and Case Study (RQ4)}
Finally, we conduct a case study to figure out what is this enhanced representation really learning. Although it is difficult to exactly describe what a high dimensional vector represents, its semantic information can still be inferred by comparing distances between vectors. If we already know a vector $a$ representing information A, and if another vector $b$ is close to $a$ in their latent space, we could tell that $b$ also represents information A to a great extent. To figure out the meaning of enhanced representation, we choose several representations for reference: \textbf{1)} historical items which have the same category as target item (we call them reference items for brevity), we label them from Item1 to Item5 as shown in figure \ref{visual}. We choose them because they have the same category as the target item, thus can be treated as representing similar semantics. \textbf{2)} user representation. \textbf{3)} target item embedding. Here we sampled two users from each testing set. For each user, the target item must have a frequency lower than 10. Then we calculate their representations mentioned above and plot the cosine distances between them in figure \ref{visual}.

First, we can focus on the 'enhanced' row and the 'item' row of each subgraph. As we can see, for most reference items their cosine distances between enhanced representation are smaller than that of target item embedding. That is to say, for low frequency items, enhanced representation can represent the semantics of the target item more accurately than its raw embedding. Second, one of our purposes of proposing enhanced representation is to alleviate the problem that target item embedding with poor quality leads to the mismatch between target item embedding and user representation. From figure \ref{visual} we can see that cosine distances between enhanced representation and user representation are smaller than that between target item embedding and user representation. In another word, compared with the row embedding of the target item, enhanced representation is more compatible with user representation, which can reduce the difficulty of learning the scoring function.

\begin{table}[t]
\caption{Comparison of time consumption in different stages. We report the time cost of each epoch on the Beauty dataset as the training time and report the average testing time of each user on Beauty dataset as the testing time.}
\label{statistic}
\centering
\label{time}
\begin{tabular}{l r r r r r r}
\toprule
Methods & MF & MLP & FISM & NAIS & DIN & CER \\
\midrule
Training (s)       & 41.67  & 58.24 & 47.78   & 83.78   & 111.92  & 173.80 \\

Testing (ms)      & 0.3& 0.6  & 0.9    & 2.2    & 3.1   & 3.3 \\

\bottomrule
\end{tabular}
\end{table}

\subsection{Time Complexity}
To intuitively reflect the time complexity of each model in the training and testing stage, we report the training time cost of each epoch and the average testing time of each user on the Beauty dataset in table \ref{time}. All experiments are conducted with the same hardware environment. In the training stage, our proposed CER cost a lot compared to other methods due to the calculation of enhanced representation. However, in the testing stage, our method shows a similar inference time compared to DIN or NAIS. It benefits from that the enhanced representation can be calculated offline after the training stage and directly used in the testing stage.

\section{Conclusion}
In this paper, we propose a novel enhanced representation of the target item and an item-based CF model based on enhanced representation named CER for the recommendation. Our key argument is that the quality of target item representation plays an important role in the performance of existing ICF methods and the long-tail distribution deteriorates the quality of the target item embedding. The proposed enhanced representation selectively learns semantic information from the co-occurrence items of target item, thus alleviates the lack of semantics caused by the long-tail distribution. Extensive experiments on two public datasets demonstrate the effectiveness of our proposed model.

\bibliographystyle{ACM-Reference-Format}
\bibliography{references}


\end{document}